\documentclass[12pt]{article}
\usepackage{amsfonts}
\usepackage{amssymb}
\usepackage{graphicx}
\usepackage{amsmath}
\usepackage{harvard}

\input{tcilatex}
\begin{document}

\title{The Utility of Reliability and Survival}
\date{Revised, July 2009}
\author{Nozer D. Singpurwalla \\
The George Washington University, Washington, D.C. 20052 }
\maketitle

\begin{abstract}
Reliability (survival analysis, to biostatisticians) is a key ingredient for
making decisions that mitigate the risk of failure. The other key ingredient
is utility. A decision theoretic framework harnesses the two, but to invoke
this framework we must distinguish between chance and probability. We
describe a functional form for the utility of chance that incorporates all
dispositions to risk, and propose a probability of choice model for
eliciting this utility. To implement the model a subject is asked to make a
series of binary choices between gambles and certainty. These choices endow
a statistical character to the problem of utility elicitation. The workings
of our approach are illustrated via a live example involving a military
planner. The material is general because it is germane to any situation
involving the valuation of chance.

\bigskip

\medskip\ Risk Analysis.
\end{abstract}

\setlength{\topmargin}{-0.2in} \addtolength{\textheight}{0.8in}

\setlength{\baselineskip}{22pt}\pagebreak \pagebreak

\section{\protect\underline{{\protect\large Introduction and Overview}}}

\subsection*{\textit{1.1 \ \textbf{Preamble: Motivation and Objectives}}}

Perhaps a better title for this paper could be "The Utility of Chance", but
it would detract from its motivating import, which is applied and pragmatic.
\ Indeed, the work here was suggested by a scenario wherein the author was
asked to determine if the reliability of an amphibious landing tank, called
the "Expeditionary Fighting Vehicle" was in excess of .999, a number
sacrosanct to a commanding general of the U.S. \ Marine Corps. By
reliability, we mean the survival function evaluated at any specified time,
called the "mission time". \ Given the vehicle's architecture, the said
number was literally impossible to achieve. \ Thus arose the question of why
.999? \ It turned out, as is usual the case with reliability specifications,
that such numbers are arbitrary, more a matter of decree than a
consideration of need. \ See, for example, The Washington Post, February 2,
2008 article on "GAO Report Criticizes Defense on its Acquisition of
Weapons". A more recent, albeit related, case in point is the U. S. Air
Force's decision to opt for the oversized and cost-ineffective KC-30 tanker
over the right-sized and less costly KC-767 tanker [cf. Washington Post,
June 11, 2008, p. A16].\ It is possible that similar situations may also
prevail in the biomedical environment wherein choices that impact patient
survival over cost and quality of life considerations, are not judiciously
balanced.

There are two aims that underlie this paper. \ The first is to advocate the
need for utility considerations in the reliability arena including a
suggestion for its general shape. \ The second is to propose a statistical
approach based on the item response theory models (also known as a \textbf{%
\textit{choice models}}) for eliciting an individual's utility. \ However,
to set a formal stage for achieving the first aim, we need to distinguish
between reliability (or the survival function) as an unknown \textbf{\textit{%
chance}} or \textbf{\textit{propensity}}, and \textbf{\textit{survivability}}
as ones subjective (personal) probability about the unknown chance. \ By
propensity we mean the purported \textbf{\textit{causes}} of observed stable
relative frequencies; see Popper (1957). \ Propensities are invoked to 
\textbf{\textit{explain why}} repeating a certain kind of experiment will
generate a given outcome type at a persistent rate; they are constrained to
be between 0 and 1, both inclusive. \ Regarding the second aim, statistical
methods for eliciting utilities are virtually non-existent, and regrettably
so, because such methodologies can vastly enhance the utility assessment
enterprise. \ Some exceptions are the papers of Mosteller and Nogee (1951),
Becker, De Groot and Marschak (1964), and Novick and Lindley (1979).

Stripped of the reliability centered application that has motivated our
work, the underlying theme of this paper should have a wider appeal. \ It is
germane to any situation that entails the desirability of a chance. \ For
example, how much more desirable is a coin with a propensity of heads of say
.95 to a coin whose propensity of heads is .93, given that the coins are to
be used for gambling? \ Similarly, how much more desirable is a pill with a
cure rate of .98 to one with a cure rate of .95, given that the former could
cost much more than the latter? \ To address issues such as these we need to
assess the utility of propensity. \ But first in order are some comments on
the roles of reliability and survival analysis in risk management, the roles
of probability and utility decision in making, and the structure of a
decision problem. \ Sections 1.2 through 1.4 are devoted to these topics. \
Section 1.5 is a summary presentation of de Finetti's (1972) theorem on
infinite exchangeable Bernoulli sequences, and how this theorem may provide
a hook on which the essence of the material here can be thought of. \ The
rest of this paper, Sections 2, 3, 4, and 5, pertains to the utility of
reliability, a model for eliciting utilities, deploying the model, and a
live application, respectively. \ Section 6 closes this paper. \ It may be
of relevance to note that the matter of a utility of chance is to be
contrasted with that of the utility of probability for which there is a
precedence in the works of Lindley (1976), and Good and Card (1971).

\subsection*{\textit{1.2 \ \textbf{Reliability and Survival Analysis in Risk
Management}}}

Reliability (Survival) analyses done by engineers (biostatisticians) provide
yardsticks for quantifying the random nature of lifetimes. \ We quantify
this randomness for managing the risk of failure. \ Managing risk means
making choices that minimize the losses caused by adverse events. \ In the
context of engineering, such choices pertain to deciding between competing
designs, managing maintenance, or the acceptance/rejection of manufactured
lots. \ In biomedicine, decision making pertains to treatment options, and
other choices that impact survival and the quality of life.

\subsection*{\textit{1.3 \ \textbf{Coherent Decision Making: Probability and
Utility}}}

Coherent decision making rests on two pillars, probability and utility, and
the principle of maximization of expected utility (MEU). \ Probability
quantifies uncertainty and utility quantifies preferences. \ Utility in
statistical inference is via loss functions, such as squared error, absolute
error, linex, etc.. \ Such loss functions are stylized. \ There appears to
be a dearth of literature in statistical outlets about eliciting the actual
loss functions of decision makers. \ This state of affairs is also true in
the engineering sciences such as filtering, control, and information fusion,
wherein a use of squared-error loss functions seems to be the norm.

\subsection*{\textit{1.4 \ \textbf{The Structure of a Decision Problem}}}

Suppose that a decision maker $\mathcal{D}$ is required to choose one among
a set of $n$ mutually exclusive and exhaustive actions, $a_{1}$, $a_{2}$%
,..., $a_{n}$. \ Associated with $a_{i}$, are $k_{i}$ outcomes (states of
nature) $\theta _{ij}$, $j=1,...,k_{i}$, assumed mutually exclusive and
exhaustive. \ When $\mathcal{D}$ chooses $a_{i}$, $\mathcal{D}$ is uncertain
about the outcome. Let $\mathbf{P}(\theta _{ij})$ be $\mathcal{D}$'s
probability of occurrence of $\theta _{ij}$, $i=1,...,n,$ $j=1,...,k_{i}$; $%
\mathbf{P}(\theta _{ij})$ is \underline{personal} to $\mathcal{D}$. \ Let $%
U(\theta _{ij})$ be $\mathcal{D}$'s utility of $\theta _{ij}$; $U(\theta
_{ij})$ is also personal to $\mathcal{D}$. \ $U(\bullet )$ is a numerical
quantity between 0 and 1, and if $U(\theta _{ij})>U(\theta _{im})$, $j\neq m$%
, then $\mathcal{D}$ prefers $\theta _{ij}$ to $\theta _{im}$. \ We assume
that all the $\theta _{ij}$'s can be preference ranked by $\mathcal{D}$
(i.e. $\mathcal{D}$ satisfies the \textbf{\textit{axiom of completeness}}).
\ The focus of this paper is to develop a procedure for obtaining $U(\theta
_{ij})$, in a manner that ensures a certain kind of consistency; this will
be clarified later in Section 4.1. The expected utility of $a_{i}$ is,%
\begin{equation*}
\mathbf{E}[U(a_{i})]=\underset{j}{\dsum }U(\theta _{ij})\mathbf{P}(\theta
_{ij})
\end{equation*}%
and the MEU principle prescribes that $\mathcal{D}$ choose that $a_{i}$ for
which $\mathbf{E}[U(a_{i})]$ is a maximum. \ In developing an argument for
this principle $\mathcal{D}$ assumes that there exists a $\theta _{ij}$, say 
$\theta ^{\ast }$ for which $U(\theta ^{\ast })=1$, and some other $\theta
_{ij}$, say $\theta _{\ast }$ for which $U(\theta _{\ast })=0$. \ The $%
\theta ^{\ast }$ and $\theta _{\ast }$ are known as \textbf{\textit{anchor
points}}. \ In our particular application the $\theta _{ij}$ are
propensities; thus $\theta _{ij}\in \lbrack 0,1]$, with anchor points $U(0)=0
$, and $U(1)=1$.

To harness the notions of reliability and survival analysis for risk
management, we need to cast them in the decision making framework described
above. \ One approach to doing so is motivated by de Finetti's Theorem
(1972) on exchangeable sequences. \ The theorem leads to the view that the
reliability and the survival function are a chance (or a propensity), and
not a probability. \ By contrast, Kolmogorov (1969) does not distinguish
between chance and probability, so that to him reliability is a probability.
\ It is not clear as to how under this conventional view of Kolmogorov,
reliability and survival functions can be formally cast in a decision
theoretic framework.

\subsection*{\textit{1.5 \ \textbf{de Finetti's Theorem: Infinite
Exchangeable Sequences}}}

Let $X_{1},X_{2},...,$ be an infinite sequence of (non-negative) continuous
random variables with the property that for some $\mathbf{\theta }$ and all $%
i$%
\begin{equation*}
\mathbf{P}(X_{i}\geq x|\mathbf{\theta })=\overline{F}(x|\mathbf{\theta }),%
\text{ }x\geq 0.
\end{equation*}%
Then for every finite $n\geq 1$,%
\begin{equation*}
\mathbf{P}(X_{1}\geq x_{1},...,X_{n}\geq x_{n})=\overset{\infty }{\underset{0%
}{\int }}\underset{1}{\overset{n}{\tprod }}\overline{F}(x_{i}|\mathbf{\theta 
})\Pi (\mathbf{\theta })d\mathbf{\theta },
\end{equation*}%
where $\Pi (\mathbf{\theta })$ encapsulates $\mathcal{D}$'s uncertainty
about $\mathbf{\theta }$. \ Thus for $n=1$%
\begin{equation}
\mathbf{P}(X\geq x)=\underset{0}{\overset{\infty }{\tint }}\overline{F}(x|%
\mathbf{\theta })\Pi (\mathbf{\theta })d\mathbf{\theta };  \tag{1.1}
\end{equation}%
$\overline{F}(x|\mathbf{\theta })$ is the reliability (or the survival
function) of $X$, and is likened to chance; see Lindley and Phillips (1976),
or Lindley and Singpurwalla (2002). The quantity $\mathbf{P}(X\geq x)$ is $%
\mathcal{D}$'s uncertainty about the event $(X\geq x)$ described via a
personal probability. \ We label it as the \textbf{\textit{survivability}}
of $X$, and is distinguished from the \textbf{\textit{survival function}} $%
\mathbf{P}(X\geq x|\mathbf{\theta })$. In a decision theoretic set-up, $%
\mathbf{P}(X\geq x|\mathbf{\theta })=\overline{F}(x|\mathbf{\theta })\in
\lbrack 0,1]$ is the unknown state of nature, and $\Pi (\mathbf{\theta })$
encapsulates $\mathcal{D}$'s uncertainty about it.

The goal of this paper is to assess the utility of $\overline{F}(x|\mathbf{%
\theta })$, for fixed $x\geq 0$.

\section{\protect\underline{{\protect\large The Utility of Reliability / The
Survival Function}}}

Let $U[\overline{F}(x|\mathbf{\theta })]$ denote $\mathcal{D}$'s utility of $%
\overline{F}(x|\mathbf{\theta })$. With $\overline{F}(x|\mathbf{\theta })\in
\lbrack 0,1]$, we anchor on two points $\overline{F}(x|\mathbf{\theta })=1$
and $\overline{F}(x|\mathbf{\theta })=0$, setting $U(1)=1$ and $U(0)=0$, and
ask what is $U(\overline{F}(x|\mathbf{\theta }))$ for any $\overline{F}(x|%
\mathbf{\theta })\in (0,1)$?

To address this question, we offer $\mathcal{D}$ the following two choices:

\qquad i) \ Receive $\overline{F}(x|\mathbf{\theta })$ for sure, or

\qquad ii) \ Agree to a gamble wherein $\mathcal{D}$ receives $\overline{F}%
(x|\mathbf{\theta })=1$ with chance $p$, and $\overline{F}(x|\mathbf{\theta }%
)=0$ with chance $(1-p)$.

Then, $\mathcal{D}$'s utility of $\overline{F}(x|\mathbf{\theta })$ is that
value of $p$ for which $D$ is indifferent between the two choices of
certainty versus uncertainty -henceforth a $\mathit{\mathbf{p}}$\textit{%
\textbf{-gamble}}. \ With $U[\overline{F}(x|\mathbf{\theta })]$ so anchored
archetypal utility functions can be prescribed by the relationship $U[%
\overline{F}(x|\mathbf{\theta })]=(\overline{F}(x|\mathbf{\theta }))^{\frac{%
\beta }{x}}$, for $\beta >0$; see Figure 2.1.\ Note that $\mathcal{D}$ is
risk neutral for $\beta =x$, and risk prone (averse) for $\beta >(<)$ $x$.

\bigskip 
\begin{equation*}
\FRAME{itbpF}{3.7706in}{2.4898in}{0pt}{}{}{Figure}{\special{language
"Scientific Word";type "GRAPHIC";maintain-aspect-ratio TRUE;display
"USEDEF";valid_file "T";width 3.7706in;height 2.4898in;depth
0pt;original-width 4.3241in;original-height 2.8452in;cropleft "0";croptop
"1";cropright "1";cropbottom "0";tempfilename
'KMJ1QG0X.wmf';tempfile-properties "XPR";}}\FRAME{itbpF}{3.6123in}{2.2675in}{%
-16.5625pt}{}{}{Figure}{\special{language "Scientific Word";type
"GRAPHIC";maintain-aspect-ratio TRUE;display "USEDEF";valid_file "T";width
3.6123in;height 2.2675in;depth -16.5625pt;original-width
3.9998in;original-height 2.5002in;cropleft "0";croptop "1";cropright
"1";cropbottom "0";tempfilename 'KMJ1QG0Y.wmf';tempfile-properties "XPR";}}
\end{equation*}%
\qquad 
\begin{equation*}
\text{\underline{\text{Figure 2.1}}: Archetypal Forms of }\mathcal{D}\text{%
's Utility of }\overline{F}(x|\mathbf{\theta })\text{.}
\end{equation*}

It can be so that $\mathcal{D}$ is risk prone for small values of $\overline{%
F}(x|\mathbf{\theta })$ and risk averse for large values of $\overline{F}(x|%
\mathbf{\theta })$ making the utility function S-shaped; similarly, a
reverse S-shape, mutatis-mutandis.

\subsection*{\textit{2.1 \ }\textbf{\textit{Eliciting Utility: Some Caveats.}%
}}

For convenience set $\overline{F}(x|\mathbf{\theta })=c$, for $c\in (0,1)$.
Recall that $U(c)$ is that chance, say $p^{\ast }$, at which $\mathcal{D}$
is indifferent between receiving a $c$ for sure, versus a $p^{\ast }$-gamble.

The conventional approach for eliciting $p^{\ast }$, revolves around two
methods, \textbf{\textit{fixed probability}}, and \textbf{\textit{fixed state%
}} [cf. Hull, Moore, and Thomas (1973), or Farquhar (1984)]. \ In the
former, $\mathcal{D}$ is presented with a $p^{\ast }$-gamble, and is asked
to choose a $c\in (0,1)$ for which $\mathcal{D}$ is indifferent between $c$
and the $p^{\ast }$-gamble. \ The $c$ so chosen is called the \textbf{%
\textit{certainty equivalent}} of the gamble. \ This exercise is repeated
for a range of values of $p^{\ast }\in (0,1)$. \ In the fixed state method, $%
c$ is fixed and $\mathcal{D}$ is interrogated over a range of values of $p$
until $\mathcal{D}$ converges on a $p^{\ast }$ for which $\mathcal{D}$ is
indifferent between receiving the fixed $c$ and a $p^{\ast }$-gamble. \
Either method is cumbersome because it is difficult to iterate around an
indifference value of $c$ or $p$. \ However, for any fixed $p$, it may be
easier for $\mathcal{D}$\ to make a binary choice between receiving a sure $%
c $ versus a $p$-gamble. \ Indeed, this is the essence of our proposed
choice model based approach for eliciting utility; this approach is
discussed in Sections 3.

In addition to the elicitation difficulty mentioned above, there are two
other issues associated with the conventional approach. \ For one, there is
no assurance that $\mathcal{D}$ will be consistent in the declared values of 
$p$. \ Specifically, the $p$'s should be non-decreasing in $\overline{F}(x|%
\mathbf{\theta })$ -the \textbf{\textit{monotonicity requirement}}- and they
must be invariant with respect to the anchor points used to elicit them -the 
\textbf{\textit{invariance requirement}}. \ Attempts at resolving this
latter type of inconsistency have entailed a use of linear programming and
regression based approaches [cf. Meyer and Pratt (1968), and Novick and
Lindley (1979), respectively].

Another feature of the conventional approach is that it being fundamentally
deterministic and iterative, there is no provision for accommodating $%
\mathcal{D}$'s lack of definitiveness about the declared values. \ By
contrast, a Bayesian procedure based on binary choices will have a built in
mechanism for the treatment of $\mathcal{D}$'s uncertainties.

Finally, there is a price to be paid for achieving a high reliability and
the survival probability. \ This tantamounts to a \textbf{\textit{disutility}%
}. \ The matter of disutilities associated with high survival probabilities,
has spawned the topic of "\textit{\textbf{quality of life}}" in the health
sciences, [cf. Mesbah and Singpurwalla (2008)].

\subsection*{\textit{2.2 \ }\textbf{\textit{Incorporating Disutility: An
Omnibus Utility.}}}

For purposes of illustration, an archetypal function that is able to reflect
the feature that an increase in reliability should be accompanied by an
increase in cost, so that the disutility due to cost is an increasing
function of reliability, can be of the form%
\begin{equation}
1-\exp \left( -\frac{\delta \overline{F}(x|\mathbf{\theta })}{1-\overline{F}%
(x|\mathbf{\theta })}\right) \text{,}  \tag{2.1}
\end{equation}%
for some $\delta >0$.

Combining this disutility with the utility, yields an \textbf{\textit{%
omnibus utility }}for reliability (or survival) as:%
\begin{equation}
(\overline{F}(x|\mathbf{\theta }))^{\frac{\beta }{x}}-\left[ 1-\exp \left( -%
\frac{\delta \overline{F}(x|\mathbf{\theta })}{1-\overline{F}(x|\mathbf{%
\theta })}\right) \right] ,  \tag{2.2}
\end{equation}%
for $x\geq 0$, $\beta ,$ $\delta >0,$ and $\overline{F}(x|\mathbf{\theta }%
)\in \lbrack 0,1]$. \ Like Equation (2.1), Equation (2.2) is also
illustrative.

\section{\protect\underline{{\protect\large A Probability of Choice Model
for Utility Elicitation}}}

The thesis that it is easier for $\mathcal{D}$ to make a binary choice
between the options of receiving an $\overline{F}(x|\mathbf{\theta })\in
(0,1)$ for sure, or receiving $\overline{F}(x|\mathbf{\theta })=1(0)$ with
chance $p(1-p)$, versus arriving upon a $p$ by iteration motivates us to
consider \textbf{\textit{Probability of Choice Models}} as a possible
mechanism for eliciting utilities. In Section 3.1 we motivate and introduce
our model. A use of this model entails fixing $c$ at some $c_{i}$, and $p$
at $p_{ij}$, and then asking $\mathcal{D}$ to make a binary choice between a
sure $c_{i}$ versus a $p_{ij}$-gamble, for $j=1,2,...,n_{i}$. \ We set $%
Y_{ij}=1(0)$ if $\mathcal{D}$ opts (does not opt) for the gamble. We repeat
this exercise for different values of $c_{i},$ $i=1,2,...,n_{i}$. \ Using
the $Y_{ij}$ as data, we estimate the model parameters, and for every $c$,
find that $p$ for which the probability of choosing the $p$-gamble is $0.5$.
\ This binary choice strategy resonates with the method used by Mosteller
and Nogee (1951) who consider the proportion of times a subject chooses the
various gambles.

\subsection*{\textit{3.1 \ }\textbf{\textit{The Proposed Model for Utility
Elicitation}}}

The proposed model is based on how hard or easy it is for $\mathcal{D}$ to
make the binary choices. \ Given below are the boundary conditions for a
rational $\mathcal{D}$:

$\bullet $ \ When $p=1$, $\mathcal{D}$ will choose $Y=1$ for all $c<1$;

$\bullet $ \ When $p=0$, $\mathcal{D}$ will choose $Y=0$ for all $c>0$;

$\bullet $ \ When $p=c$, a \textit{\textbf{risk neutral}} $\mathcal{D}$ will
choose $Y=1$ or $Y=0$, equally often; by

contrast, a \textit{\textbf{risk prone}} (\textbf{\textit{averse}}) $%
\mathcal{D}$ will choose $Y=1$ more (less) often than $Y=0.$

Let $\mathbf{P}(Y=1)=\Pi $ denote an elicitor $\mathcal{E}$'s personal
probability of $\mathcal{D}$ choosing a $p$-gamble over the certain $c$. \
Note that there are two kinds of entities that come into play; $\mathcal{E}$%
's personal probability $\Pi $, and a chance $p$. \ $\mathcal{D}$'s
indifference between the two choices tantamounts to $\Pi =\frac{1}{2}$. \ $%
\mathcal{D}$'s choices are the easiest with $\mathcal{E}$'s $\Pi =1$, when $c
$ and $p$ lie on the lines joining $(c=0,p=0)$, $(0,1]$, and $(1,1)$ and
with $\Pi =0$, when they lie on the lines joining $(0,0)$, $(1,0]$ and $(1,1)
$. \ These are the boundaries of Figure 3.1. \ $\mathcal{D}$'s choices
become difficult as $p$ and $c$ get close to each other, becoming the most
difficult when $c=p$. \ The roles of $p$ and $c$ are analogous to those of
the ability and the difficulty parameters in the Rasch Model used in
education testing and quality of life studies [cf. Mesbah, Cole and Lee
(2002)].\pagebreak 

\FRAME{dtbpF}{4.1753in}{2.5417in}{0pt}{}{}{Figure}{\special{language
"Scientific Word";type "GRAPHIC";maintain-aspect-ratio TRUE;display
"USEDEF";valid_file "T";width 4.1753in;height 2.5417in;depth
0pt;original-width 4.1252in;original-height 2.5002in;cropleft "0";croptop
"1";cropright "1";cropbottom "0";tempfilename
'KMJ1QG0Z.wmf';tempfile-properties "XPR";}}

\begin{equation*}
\text{\underline{Figure 3.1}: Boundary Conditions for a Risk Neutral }%
\mathcal{D}
\end{equation*}

The boundary conditions of Figure 3.1 motivate the general forms of Figure
3.2 as $\mathcal{E}$'s model for $\Pi $, seen as a function of $(p-c)$.%
\FRAME{dtbpF}{4.804in}{3.0441in}{0pt}{}{}{Figure}{\special{language
"Scientific Word";type "GRAPHIC";maintain-aspect-ratio TRUE;display
"USEDEF";valid_file "T";width 4.804in;height 3.0441in;depth
0pt;original-width 4.7504in;original-height 3in;cropleft "0";croptop
"1";cropright "1";cropbottom "0";tempfilename
'KMJ1QG10.wmf';tempfile-properties "XPR";}}

\begin{equation*}
\text{\underline{Figure 3.2}: General Forms for }\mathcal{E}\text{'s Model
for }\Pi \text{.}
\end{equation*}

\bigskip

For a risk neutral $\mathcal{D}$, the right diagonal line of Figure 3.2
would encapsulate $\mathcal{E}$'s choice probabilities, whereas the concave
(convex) curve would encapsulate these probabilities for a risk prone
(averse) $\mathcal{D}$. \ The above relationships between $(p-c)$ and $\Pi $
can, for $p,c\neq 0,1$, be encapsulated via the equation%
\begin{equation}
\Pi =\mathbf{P}(Y=1|\beta ;c,p)=\left( \frac{(p-c)+1}{2}\right) ^{\beta }. 
\tag{3.1}
\end{equation}

The inclusion of an additional parameter $\alpha >0$ enables us to
incorporate varying degrees of risk aversion and proneness; see Figures 3.3
a) and b). \ Thus a penultimate version of the model is%
\begin{equation}
\Pi =\mathbf{P}(Y=1|\alpha ,\beta ;c,p)=\left( \frac{(p-c)^{\alpha }+1}{2}%
\right) ^{\beta }\text{.}  \tag{3.2}
\end{equation}

Figure 3.3 a) pertains to the case of $\alpha =\frac{1}{3}$ with $\beta =%
\frac{1}{2}$, $1$, and $2$ respectively, and Figure 3.3 b) is for $\alpha =3$
with similar values for $\beta $. 
\begin{equation*}
\FRAME{itbpF}{3.4212in}{2.668in}{0in}{}{}{Figure}{\special{language
"Scientific Word";type "GRAPHIC";maintain-aspect-ratio TRUE;display
"USEDEF";valid_file "T";width 3.4212in;height 2.668in;depth
0in;original-width 3.3754in;original-height 2.6247in;cropleft "0";croptop
"1";cropright "1";cropbottom "0";tempfilename
'KMJ1QG11.wmf';tempfile-properties "XPR";}}\FRAME{itbpF}{3.4212in}{2.668in}{%
0in}{}{}{Figure}{\special{language "Scientific Word";type
"GRAPHIC";maintain-aspect-ratio TRUE;display "USEDEF";valid_file "T";width
3.4212in;height 2.668in;depth 0in;original-width 3.3754in;original-height
2.6247in;cropleft "0";croptop "1";cropright "1";cropbottom "0";tempfilename
'KMJ1QG12.wmf';tempfile-properties "XPR";}}
\end{equation*}

\begin{equation*}
\text{\underline{Figure 3.3}: Illustration of }\mathcal{D}\text{'s
Probability of Choice.}
\end{equation*}

Whereas Equation (3.2) is intuitively appealing, it suffers from a technical
deficiency which arises when we wish to solve for $(p-c)$ with $\Pi $ set at 
$0.5$. \ Specifically, to avoid complex roots, only certain combinations of
values of $\alpha $ and $\beta $ are admissible.\ Thus as a refinement of
Equation (3.2), the final version of our model for $\mathcal{E}$'s choice
probability is:%
\begin{equation}
\mathbf{P}(Y=1|\alpha ,\beta ;c,p)=\left\{ 
\begin{array}{c}
0,\text{ \ \ \ \ \ \ \ \ \ \ if }p=1\text{ and }c<1\text{, or }p>1\text{ and 
}c=0\text{;\bigskip } \\ 
\frac{1}{2},\text{ \ \ \ \ \ \ \ \ \ \ if }p=0\text{ and }c=0\text{, or }p=1%
\text{ and }c=1\text{;\bigskip } \\ 
1,\text{ \ \ \ \ \ \ \ \ \ \ if }p=0\text{ and }c>0\text{, or }p<1\text{ and 
}c=1\text{;\bigskip } \\ 
\frac{1}{2}\left[ 1+\text{sgn}(p-c)|p-c|^{\alpha }\right] ^{\beta },\text{ \
\ \ \ \ \ \ \ \ \ \ \ \ \ \ \ \ otherwise,}%
\end{array}%
\right.  \tag{3.3}
\end{equation}%
where sgn$(\mathfrak{z})=-1(+1)[0]$, when $\mathfrak{z}<(>)[=]0$, and $%
\alpha ,\beta >0$.

\section{\protect\underline{{\protect\large Implementing the Model}}}

For any fixed $c$, our aim is to find that $(p-c)$ for which $\mathbf{P}%
(Y=1|\alpha ,\beta ;c,p)=\frac{1}{2}$; we can then solve for $p$, $\mathcal{D%
}$'s $p$-gamble, for the fixed $c$.\ To do the above we need estimates of $%
\mathcal{D}$'s values of $\alpha $ and $\beta $ for the chosen $c$. This can
be done using the $Y_{ij}$'s declared by $\mathcal{D}$, for $c$ fixed at $%
c_{i}$, and a range of values of $p$, say $p_{ij}$, $j=1,...,n_{i}$.

In actual practice the values chosen for $c_{i}$ and $p_{ij}$ will not
involve the boundary conditions since there is no need to elicit preferences
for such clear cut situations. \ For purposes of discussion, suppose that
the data $Y_{ij}$, $j=1,...,n_{i}$, yield $\widehat{\alpha }_{i}$ and $%
\widehat{\beta }_{i}$ \ as the maximum likelihood estimators of $\alpha $
and $\beta $, respectively; see Appendix A. \ Then, the desired $%
(p_{i}-c_{i})\overset{def}{=}\omega _{i}$, which is a solution to the
relationship $\mathbf{P}(Y_{i}=1|\widehat{\alpha }_{i},\widehat{\beta }%
_{i};c_{i},p_{i})=\frac{1}{2}$, will be of the form:%
\begin{equation}
\widehat{\omega }_{i}=\text{sgn}(\widehat{\beta }_{i}-1)\left[ \text{sgn}(%
\widehat{\beta }_{i}-1)(2^{1-\frac{1}{\widehat{\beta }_{i}}}-1)\right] ^{%
\frac{1}{\widehat{\alpha }_{i}}};  \tag{4.1}
\end{equation}%
also, $\widehat{\omega }_{i}\in \lbrack -1,+1]$.

Consequently, with utilities being constrained to lie between $0$ and $1$, $%
\mathcal{D}$'s utility for $c_{i}$ will be:%
\begin{equation}
U(c_{i})=\left\{ 
\begin{array}{c}
\min (1,c_{i}+\widehat{\omega }_{i}),\text{ \ \ \ \ \ if }\widehat{\omega }%
_{i}>0\text{,\bigskip } \\ 
\max (0,c_{i}+\widehat{\omega }_{i}),\text{ \ \ \ \ \ if }\widehat{\omega }%
_{i}<0\text{,\bigskip } \\ 
c_{i},\text{ \ \ \ \ \ \ \ \ \ \ \ \ \ \ \ \ \ \ \ \ \ \ \ \ \ if }\widehat{%
\omega }_{i}=0\text{.}%
\end{array}%
\right.  \tag{4.2}
\end{equation}

When $\widehat{\omega }_{i}<(>)[=]0$, $\mathcal{D}$ is risk prone (averse)
[neutral] for $c_{i}$. \ The above exercise is repeated for a range of
values of $c$, say $c_{1},...,c_{i},...,c_{n}$, to yield $\mathcal{D}$'s
utility for survival as $U(c_{1}),...,U(c_{n})$, as perceived by $\mathcal{E}
$.

\subsection*{\textit{4.1 \ Adjacent Point Gambles and Coherence}}

The elicitation of $\mathcal{D}$'s $Y_{ij}$'s discussed in Section 3.2
presumes \textbf{\textit{end point gambles}}; i.e. $\mathcal{D}$ receives
either a sure $c_{i}$, or a $1$ with chance $p_{ij}$ and a $0$ with chance $%
(1-p_{ij})$. \ A problematic feature of these end point gambles is that the
resulting utilities may not be increasing (non-decreasing) in $c_{i}$. \
This obstacle can be tempered by considering \textit{\textbf{adjacent point
gambles}}. \ That is, $\mathcal{D}$ is offered a choice between a sure $%
c_{i} $ and a $p_{ij}$ gamble involving $U(c_{i-1})$ and $U(c_{i+1})$, with $%
U(c_{0})=U(0)=1$, and $U(c_{n+1})=U(1)=1$, $i=1,...,n$.

The adjacent gamble process can start-off with some $c_{i}$ and an end point
gamble to obtain a $U(c_{i})$. \ This is followed by picking a $c_{k}$
between $c_{i}$ and $1$ (or between $0$ and $c_{i}$) to obtain a $U(c_{k})$
via a $p_{kj}$-gamble involving $U(c_{i})$ and $1$ (or involving $0$ and $%
U(c_{i})$). \ Once $U(c_{i})$ and $U(c_{k})$ are at hand, these can be used
to obtain $U(c_{m})$ for $c_{i}<c_{m}<c_{k}$ via a sure $c_{m}$ versus a $%
p_{mj}$-gamble.

The adjacent gamble approach will help achieve the monotonicity requirement
of the utility function but will not guarantee it. \ This is because of the
inherent randomness in the proposed approach which is based on estimating $%
\alpha $ and $\beta $. \ For a $c_{m}\in (c_{i},c_{k})\,$, should $U(c_{m})$
be greater (less) than $c_{k}$ ($c_{i}$), then a way to achieve monotonicity
would be to let $U(c_{m})$ lie on the line segment joining $U(c_{i})$ and $%
U(c_{k})$. A strategy such as this is used in \textbf{\textit{isotonic
regression}} [see, for example, Barlow, Bartholomew, Bremner, and Brunk
(1972)].

The matter of resolving inconsistencies caused by the failure of invariance
is more difficult to deal with. \ Novick and Lindley (1979) propose
elicitation based on \underline{all} possible triplets of the form $0\leq
i<m<k\leq 1$, each triplet entailing a gamble, which we denote as a $p_{imk}$%
-gamble. \ A least-squares analysis involving the minimization of the $%
U(\bullet )$'s with respect to log-odds of the type%
\begin{equation*}
\underset{i,m,k}{\sum }\left[ \log (p_{imk}/1-p_{imk})-\log
(U(c_{m})-U(c_{i})/U(c_{k})-U(c_{m}))\right] ^{2}
\end{equation*}%
is then performed; this results in a utility function $U(c_{m})$. \ Some
discussion on an approach such as this is in Becker, De Groot and Marschak
(1963).

\subsection*{\textit{4.2 \ The Incorporation of Uncertainties: A Bayesian
Approach}}

Philosophical considerations aside, an approach based on the maximum
likelihood estimators $\widehat{\alpha }$ and $\widehat{\beta }$ suffers
from the drawback that exact measures of uncertainty about the inferred
utility function $U(\bullet )$ are difficult to obtain. \ This is so even at
the level of pointwise confidence limits for the $U(c_{i})$'s, $i=1,...,n$.
\ On the other hand, a parametric Bayesian approach is able to account for
the underlying uncertainties by the process of averaging out with respect to
the posterior distributions of $\alpha $ and $\beta $. \ To see how, suppose
that $\Pi (\alpha _{i},\beta _{i})$ denotes the joint prior distribution on $%
\alpha _{i}$ and $\beta _{i}$, for $i=1,...,n$; a specific case is described
in Section A.1 of Appendix A. \ For any fixed $c_{i}$ and data $Y_{ij}$, $%
j=1,...,n_{i}$, the above prior will lead to a posterior distribution of $%
\alpha _{i}$ and $\beta _{i}$, say $\Pi (\alpha _{i},\beta _{i};\bullet )$.
\ Then $\widetilde{\omega }_{i}$, a Bayes assessment of $\omega
_{i}=(p_{i}-c_{i})$, would be the solution of%
\begin{equation}
\underset{\alpha _{i}}{\int }\underset{\beta _{i}}{\int }\left[ \left( \frac{%
1+\text{sgn}(\omega _{i})|\omega _{i}|^{\alpha _{i}}}{2}\right) ^{\beta
_{i}}-\frac{1}{2}\right] \Pi (\alpha _{i},\beta _{i};\bullet )d\alpha
_{i}d\beta _{i}=0\text{.}  \tag{4.3}
\end{equation}

With $\widetilde{\omega }_{i}$, $i=1,...,n$ at hand, we may obtain the $%
U(c_{i})$, $i=1,...,n$, by mimicking the procedure outlined before the $%
\widehat{\omega }_{i}$ obtained via the method of maximum likelihood.

\section{\protect\underline{{\protect\large Application: The Utility of a
Vehicle's Reliability}}}

The following real life example pertains to assessing the utility of
reliability of a yet to be designed manned ground combat vehicle, as
perceived by $\mathcal{D}$, a military analyst. \ $\mathcal{D}$ is also an
officer in uniform (retired) and is thus knowledgeable about strategic
needs. \ Furthermore, $\mathcal{D}$ is a well trained operations research
analyst exposed to analytical thinking and decision making. \ Knowing the
utility of reliability (for a mission time that is specified by $\mathcal{D}$%
) will help the government procurers specify the vehicle's \textit{\textbf{%
design reliability}}. \ The vehicle in question belongs to the family of
systems called "Future Combat Systems". \ There are several hundred such new
vehicles that are to be commissioned in a brigade, each costing several
hundred thousand U.S. dollars. \ We assume exchangeability of all the new
vehicles in the brigade. \ The reliabilities of interest to $\mathcal{D}$
are in the range of 0.5 to 0.9, the lower reliabilities of no value, and the
higher reliabilities deemed unnecessary. \ Because of security
classification, no additional details can be made available.

Elicitations from $\mathcal{D}$ entail both end point gambles and adjacent
point gambles. \ Here $c_{i}$ takes values .5, .6, .7, .8 and .9, for $%
i=1,...,5$ respectively. The values chosen for the $p_{ij}$'s are .3, .4,
.5, .6, .7, .8, .9 and .95, for $j=1,...,8$, respectively, in the case of
the end point gambles, and .3, .4, .45, .5, .55, .6, and .7, for the
adjacent point gambles. \ The gamble probabilities $p_{ij}$ were chosen to
ensure that neither the gamble nor the sure thing would always be preferred
for any of the reliabilities. \ 

\FRAME{dtbpF}{4.3016in}{2.0903in}{0pt}{}{}{Figure}{\special{language
"Scientific Word";type "GRAPHIC";maintain-aspect-ratio TRUE;display
"USEDEF";valid_file "T";width 4.3016in;height 2.0903in;depth
0pt;original-width 4.2497in;original-height 2.0513in;cropleft "0";croptop
"1";cropright "1";cropbottom "0";tempfilename
'KMJ1QG13.wmf';tempfile-properties "XPR";}}

\begin{equation*}
\text{\underline{Table 5.1}: }\mathcal{D}\text{'s Choices Under End Point
Gambles.}
\end{equation*}

Table 5.1 shows the results of the elicitation for the end point gambles,
and Table 5.2\ \ for the adjacent point gambles. The entries in these tables
gives the values $Y_{ij}$, with $Y_{ij}=1$, whenever $\mathcal{D}$ opts for
the gamble. $\mathcal{D}$'s choices are solely based on strategic needs, not
costs.

An inspection of the entries in Table 5.1 suggests that $\mathcal{D}$ tends
to be risk averse, because $\mathcal{D}$ opts for the $p_{ij}$ - gamble only
when $p_{ij}\geq c_{i}$. \ The entries of Table 5.1 when used to obtain the $%
U(c_{i})$'s via the method of maximum likelihood and also the Bayesian
approach (using the independent gamma priors described in Section A.1 of
Appendix ) yields results that are almost identical; specifically $U(.5)=.5,$
$U(.6)=.6,$ $U(.7)=.7,$ $U(.8)=.93,$ $U(.9)=.92$. \ Figure 5.1 shows a plot
of the $U(c_{i})$'s versus $c_{i},$ $i=1,...,5$.\FRAME{dtbpF}{6.3114in}{%
3.5466in}{0pt}{}{}{Figure}{\special{language "Scientific Word";type
"GRAPHIC";maintain-aspect-ratio TRUE;display "USEDEF";valid_file "T";width
6.3114in;height 3.5466in;depth 0pt;original-width 6.25in;original-height
3.4999in;cropleft "0";croptop "1";cropright "1";cropbottom "0";tempfilename
'KMJ1QG14.wmf';tempfile-properties "XPR";}}%
\begin{equation*}
\text{\underline{Figure 5.1}: }\mathcal{D}\text{'s Utility of Reliability
Based on End Point Gambles (MLE and Bayes).}
\end{equation*}

Figure 5.1 suggests that $\mathcal{D}$'s utility, based on end point
gambles, is linear in $c_{i}$, save for an upward jump at .8 followed by a
slight drop of .01 at .9. This suggests that $\mathcal{D}$ is risk neutral
for values of reliability up to .7 and is risk averse for values of
reliability greater than .7. The drop in utility at .9 is an aberration
that, hopefully, can be avoided by using adjacent point gambles. The essence
of the message of Figure 5.1 is that there does not appear to be any gain in
utility in going from a reliability of .8 to a reliability of .9. Thus to
this $\mathcal{D}$, the strategic worthiness of the vehicle matures at a
reliability of about .8; higher reliabilities are of little strategic
consequence. A similar conclusion seems to be true with a consideration of
adjacent point gambles, the data for which are given in Table 5.2 below.%
\FRAME{dtbpF}{3.4428in}{1.9346in}{0pt}{}{}{Figure}{\special{language
"Scientific Word";type "GRAPHIC";maintain-aspect-ratio TRUE;display
"USEDEF";valid_file "T";width 3.4428in;height 1.9346in;depth
0pt;original-width 3.397in;original-height 1.8965in;cropleft "0";croptop
"1";cropright "1";cropbottom "0";tempfilename
'KMJ1QG15.wmf';tempfile-properties "XPR";}}%
\begin{equation*}
\text{\underline{Table 5.2}: }\mathcal{D}\text{'s Choices Under Adjacent
Point Gambles}.
\end{equation*}

An examination of the entries in Table 5.2 suggests that under adjacent
point gambles, $\mathcal{D}$ tends to be risk prone in disposition towards
the vehicle reliability for the values $c_{i}$ considered here. \ $\mathcal{D%
}$'s shift from the risk aversion phenomenon of the entries of table 5.1 to
the proneness phenomenon of Table 5.2 is intriguing. It could be attributed
to the feature that it may be easier for $\mathcal{D}$ to contemplate end
point gambles than adjacent point gambles.

\FRAME{dtbpF}{3.5665in}{2.0487in}{0pt}{}{}{Figure}{\special{language
"Scientific Word";type "GRAPHIC";maintain-aspect-ratio TRUE;display
"USEDEF";valid_file "T";width 3.5665in;height 2.0487in;depth
0pt;original-width 3.5198in;original-height 2.0098in;cropleft "0";croptop
"1";cropright "1";cropbottom "0";tempfilename
'KMJ1QH16.wmf';tempfile-properties "XPR";}}%
\begin{equation*}
\text{\underline{Table 5.3}: }\mathcal{D}\text{'s Utility of Reliability
Based on Adjacent Point Gambles}.
\end{equation*}

The data of Table 5.2 was used to obtain $\mathcal{D}$'s utility of
reliability via both the method of maximum likelihood and the Bayesian
approach (described in Section A.1 (involving independent gamma priors). \
Table 5.3 shows the results. Plots of the $U(c_{i})$ versus $c_{i}$ obtained
via the two methods are shown in Figure 5.2.

\FRAME{dtbpF}{6.3114in}{3.5466in}{0pt}{}{}{Figure}{\special{language
"Scientific Word";type "GRAPHIC";maintain-aspect-ratio TRUE;display
"USEDEF";valid_file "T";width 6.3114in;height 3.5466in;depth
0pt;original-width 6.25in;original-height 3.4999in;cropleft "0";croptop
"1";cropright "1";cropbottom "0";tempfilename
'KMJ1QH17.wmf';tempfile-properties "XPR";}}

\begin{equation*}
\text{\underline{Figure 5.2}: A Plot of }\mathcal{D}\text{'s Utility of
Reliability Based on Adjacent Point Gambles}.
\end{equation*}

The plots of Figure 5.2 suggests that for the priors chosen, the method of
maximum likelihood yields uniformly higher values for the utility function
than those yielded by the Bayesian approach, at least for the chosen priors.
\ The differences however, are not substantial, and especially so,
considering the fact that the Bayesian approach has a built in mechanism for
incorporating the underlying uncertainties. Even though the raw entries of
Table 5.2 suggest that $\mathcal{D}$ appears to be risk prone, the S-shaped
nature of the plots of Figure 5.2 suggests that $\mathcal{D}$ tends to be
risk neutral. The disparity between the intuitive conclusions that are
formed by an inspection of the raw data, and those revealed by a formal
analysis of these data is due to the smoothing of the data and the
consistency that is enforced by the model of Equations (3.3) and (4.2). \ 

Finally, the likes of Figure 5.2 suggest that there is little by the way of
utility for reliabilities .5 or below, and that there is not much, if any,
to be gained in going from a reliability of .8 to .9, at least under a
Bayesian approach. Thus the need to push for reliability numbers such as
.999 that General Officers usually tend to demand is unwarranted. Whereas we
have not performed a reliability analysis of the vehicle in question, we
have been told by $\mathcal{D}$ that the new vehicle reliability
specifications tend to be in the form of 0 failures in 1000 operational
hours, a requirement that is "fanciful and without connection to the real
world".

\section{\protect\underline{{\protect\large Summary and Conclusions.}}}

In this paper, we have advocated the point of view that a purpose for
assessing the reliability and the survival functions is to help make
decisions that mitigate risk. \ We have then cast the matter in a decision
theoretic framework by leaning on a distinction between chance and
probability. \ This has then been followed by proposing a general
architecture for the utility of chance or reliability, including some
boundary conditions. \ We have then proposed a statistical approach for
assessing utilities based on binary choices between gambles and certainties.
\ To facilitate this approach we have proposed a new choice (or item
response theory) model that has features which parallel the Rasch Model in
the sense that both models are indexed by a difference $-$ in our case $%
(p-c) $. Here, this difference is germane because the ease with which $%
\mathcal{D}$ can make a choice depends on the closeness of $c$ and $p$. The 
\textit{Grade of Membership Model}\textbf{\ }(GOM), discussed by Ershova,
Fienberg and Joutard (2007), encompass aspects of the Rasch Model [cf.
Ershova (2005)], and offers prospects for developing utility elicitation
models more sophisticated than ours. The GOM model is difficult to
appreciate and will require much thought to put it to work. \ Even though
the material of Section 3 on using a Choice Model for utility elicitation is
focused on the utility of reliability, the ideas therein are general enough
for eliciting utilities in contexts that go beyond the special case of
reliability.

There is much literature on utility theory and utility elicitation by
economists, decision theorists, game theorists, mathematicians and
statisticians. \ The names associated with these literatures are
distinguished. \ Whereas we have endeavoured to gain an appreciation of as
much of these works as is possible, it is likely that we may have missed
some contributions that make our approach and our model for utility
elicitation not new. \ But assuming that the above is not true, the work
described here may open up avenues for future investigations. \ 

\section*{\protect\underline{{\protect\large Acknowledgements}}}

Philip Wilson contributed to the development of the final elicitation model,
and the numerical work in connection with the model. \ Professor Mounir
Mesbah of University of Paris VI suggested the possibility of using the
Rasch Model for eliciting utilities. \ The military problem was brought to
our attention by Arthur Fries of the Institute for Defense Analysis (IDA).
Robert Holcomb of the IDA served as the military planner whose utilities
were assessed in Section 5. Stephen Fienberg drew our attention to the GOM
model and in so doing has opened our eyes to things that we were unaware of.
\ The comments of an insightful referee helped clarify some muddled thinking
of the author; thank you! \ The author's research has been supported by the
Office of Naval Research Contract N00014-06-1-0037 and The Army Research
Office Grants W911NF-05-1-0209 and W911NF-09-1-0039 with The George
Washington University.

\newpage

\section*{\protect\underline{{\protect\large Appendix A}}}

\bigskip

\subsection*{\textit{A.1 \ The Likelihood Function and Maximum Likelihood}}

With respect to the notation of Section 3.1, with $c$ fixed at $c_{i}$, and $%
p$ at $p_{ij}$, the likelihood of $\alpha _{i}$ and $\beta _{i}$ given the
data $y_{ij}$, $j=1,...,n_{i}$, with $y_{ij}=1$ or $0$, is of the form:%
\begin{equation}
\underset{j=1}{\overset{n_{i}}{\tprod }}\left( \frac{1}{2}\left[ 1+\text{sgn}%
(p_{ij}-c_{i})|p_{ij}-c_{i}|^{\alpha _{i}}\right] \right) ^{\beta
_{i}y_{ij}}\left( 1-\left( \frac{1}{2}\left[ 1+\text{sgn}%
(p_{ij}-c_{i})|p_{ij}-c_{i}|^{\alpha _{i}}\right] \right) ^{\beta
_{i}}\right) ^{1-y_{ij}}.  \tag{A.1}
\end{equation}

In writing out the above, we assume that given $\alpha _{i}$ and $\beta _{i}$%
, $D$'s choices for the $Y_{ij}$'s, $j=1,...,n_{i}$ are independent over the 
$p_{ij}$'s. \ This tantamounts to assuming that in making a choice $y_{ij}$, 
$\mathcal{D}$ forgets his (her) previous choices $y_{ik}$, $%
k=j-1,j-2,...,2,1 $. \ One way to achieve this independence would be to
select the $p_{ij}$ in a random order with respect to the $j$'s. \ By
avoiding choosing $c_{i}=1$ or $0$, and $p_{ij}=1$ or $0$, in the
elicitation process, we can ensure that the likelihood will not involve the
boundary conditions of Figure 3.2. \ Equation (A.1) can be maximized
numerically to yield $\widehat{\alpha }_{i} $ and $\widehat{\beta }_{i}$ as
the maximum likelihood estimators of $\alpha _{i}$ and $\beta _{i}$,
respectively.

\bigskip

\subsection*{\textit{A.2 \ Bayes Inference for }$\protect\alpha $\textit{\
and }$\protect\beta $}

An examination of Figure 3.4 shows that a large value of $\alpha $ causes $%
\Pi $ to be steeper for $(p-c)$ in the vicinity of $-1$ or $+1$, and flatter
for $(p-c)$ close to zero, than a small value of $\alpha $. \ This type of
change characterizes a $\mathcal{D}$ who switches slowly from preferring a
certain outcome to the gamble.

Such a $\mathcal{D}$ exhibits a poorer ability to discriminate between
gambles that $\mathcal{D}$ considers to be worse than the certain outcome
than those gambles that are better. \ Thus $\alpha $ may be viewed as a
parameter that encapsulates $\mathcal{D}$'s \textbf{\textit{ability to
discriminate between gambles}}, with higher values of $\alpha $ representing
a lower ability to discriminate.

Similarly, an examination of Figure 3.4 a) or b) shows that the parameter $%
\beta $ encapsulates $\mathcal{D}$'s attitude to risk, $\beta <(=)>1$
representing a risk prone (neutral) averse $\mathcal{D}$.

It is reasonable to suppose that $\mathcal{D}$'s ability to discriminate
between gambles is independent of $\mathcal{D}$'s disposition towards risk.
\ Thus $\alpha $ and $\beta $ can be treated as being independent. \ With $%
\alpha $, $\beta >0$, it is reasonable to suppose that $\alpha \lbrack \beta
]$ has a gamma distribution with scale $l[s]$ and shape $k[r]$. \ Choosing $%
k=r=2$ and $l=s=.5$, $\Pi (\alpha ,\beta )$ the joint prior on $\alpha $, $%
\beta $ is of the form $\Pi (\alpha ,\beta )\propto \beta \alpha \exp
(-2(\alpha +\beta ))$. \ Consequently, $\Pi (\alpha _{i},\beta _{i})\propto
\beta _{i}\alpha _{i}\exp (-2(\alpha _{i}+\beta _{i}))$.

Combining $\Pi (\alpha _{i},\beta _{i})$ with the likelihood (A.1) gives us
the posterior distribution of $\alpha _{i}$, $\beta _{i}$, denoted $\Pi
(\alpha _{i},\beta _{i}|\mathbf{y},\mathbf{p}_{i},c_{i})$, where $\mathbf{y}%
=(y_{i1},...,y_{i,n_{i}})$ and $\mathbf{p}_{i}=(p_{i1},...,p_{i,n_{i}})$. \
The ingredients necessary to solve Equation (4.4) are at hand, with $%
(0,\infty )$ as the limits of integration. \ The integration has been done
numerically using the bisection method.

\bigskip \pagebreak

\section*{\protect\underline{{\protect\large References}}}

\bigskip

Barlow, R. E., D. J. Bartholomew, J. M. Bremner, and H. D. Brunk (1972). 
\textit{Statistical Inference under Order Restrictions: The Theory and
Application of Isotonic Regression}. Wiley, New York.

\bigskip

Becker, G. M., M. H. DeGroot, and J. Marshak (1963). `Stochastic Models of
Choice Behavior.' \textit{Behavioral Science}, 9, 41-55.

\bigskip

Becker, G. M., M. H. DeGroot, and J. Marshak (1964). `Measuring Utility by a
Single-Response Sequential Method.' \textit{Behavioral Science}, 9, 226-232.

\bigskip

de Finetti, B. (1972). \textit{Probability, Induction and Statistics}.
Wiley, New York.

\bigskip

Ershova, E. (2005). `Comparing Latent Structures of the Grade of Membership,
Rasch, and Latent Class Models.' \textit{Psychometrica,} 70, 4, 619-628.

\bigskip

Ershova, E., Fienberg, S., and Joutard, C. (2007). `Describing Disability
Through Individual-Level Mixture Models for Multivariate Binary Data.' 
\textit{The Annals of Applied Statistics}, 1, 2, 502-537.

\bigskip

Farquhar, P. H. (1984). `Utility Assessment Methods.' \textit{Management
Science}, 30 (11): 1283-1300.

\bigskip

Good, I. J., and W. I. Card (1971). 'The Diagnostic Process with Special
Reference to Errors.' \textit{Methods of Information in Medicine,} 10, 3,
176-188.

\bigskip

Hull, J., P. G. Moore and H. Thomas (1973). `Utility and its Measurement.' 
\textit{Journal of the Royal Statistical Society A} 136 (2): 226-247.

\bigskip

Kolmogorov, A. N. (1969). `The Theory of Probability.' \textit{Mathematics,
Its Content, Methods and Meaning} (ed. Aleksandrov, A. D., Kolmogorov, A.
N., Lavrent'ev, M. A.), 2, 3, 229-264. MIT Press, Cambridge, MA.

\bigskip

Lindley, D. V. (1976). `A Class of Utility Functions.' \textit{The Annals of
Statistics}, 4, 1, 1-10.

\bigskip

Lindley, D. V., and M. R. Novick (1979). `Fixed-State Assessment of Utility
Functions.' \textit{Journal of the American Statistical Association},
306-311.

\bigskip

Lindley, D. V., and L. D. Phillips (1976). `Inference for a Bernoulli
Process (A Bayesian View).' \textit{The American Statistician}, 30, 112-119.

\bigskip

Lindley, D. V., and N. D. Singpurwalla (2002). `On Exchangeable, Causal and
Cascading Failures.' \textit{Statistical Science}, 17, 2, 209-219.

\bigskip

Mesbah, M., B. F. Cole and M. T. Lee (2002), Editors. \textit{Statistical
Methods for Quality of Life Studies}. Kluwer Academic Publishers, Boston.

\bigskip

Mesbah, M., and N. D. Singpurwalla (2008). `A Bayesian Ponders the Quality
of Life.' \textit{Statistical Models and Methods for Biomedical and
Technical Systems}, 373-384 (F. Vonta, Editor). To appear.

\bigskip

Meyer, R. F., and J. W. Pratt (1968). `The Consistent Assessment and Fairing
of Preference Functions.' \textit{IEEE Transactions on Systems Science and
Cybernetics}, SSC-4, 3, 270-278.

\bigskip

Mosteller, F., and P. Nogee (1951). `An Experimental Measurement of
Utility.' \textit{Journal of Political Economy}, 59: 371-404.

\bigskip

Popper, K. (1957). `The Propensity Interpretation of the Calculus of
Probability and of the Quantum Theory.' \textit{In Observation and
Interpretation} (Korner and Price, eds.), Bultersworth Scientific
Publications, 65-70.

\end{document}